\documentclass[review,number,sort&compress]{elsarticle}
\usepackage{lineno}

\usepackage{amssymb}
\usepackage[ngerman,english]{babel}
\usepackage{lscape}
\usepackage{rotating}

\journal{Nuclear Physics A}

\begin{document}

\begin{frontmatter}



\title{Characterization of EASIROC as Front-End for the readout of the SiPM at the focal plane of the Cherenkov telescope ASTRI }


\author[label1]{D. Impiombato}
\ead{Domenico.Impiombato@iasf-palermo.inaf.it}
\author[label1]{S. Giarrusso}
\ead{Giarrusso@iasf-palermo.inaf.it}
\author[label1]{T. Mineo}
\ead{Mineo@iasf-palermo.inaf.it}
\author[label2]{M. Belluso}
\author[label2]{S. Billotta}
\author[label2]{G. Bonanno}
\author[label1]{O. Catalano}
\author[label2]{A. Grillo}
\author[label1]{G. La Rosa}
\author[label2]{D. Marano}
\author[label1] {G. Sottile}

 \address[label1]{INAF, Istituto di Astrofisica Spaziale e Fisica cosmica di Palermo, via U. La Malfa 153, I-90146 Palermo, Italy}
 \address[label2]{INAF, Osservatorio Astrofisico di Catania, via S. Sofia 78, I-95123 Catania, Italy}


\begin{abstract}
The Extended Analogue Silicon Photo-multiplier Integrated Read Out Chip, EASIROC, is a chip
proposed as front-end of the camera at the focal plane of the imaging Cherenkov ASTRI SST-2M 
telescope prototype.
This paper presents the results of the measurements performed to characterize EASIROC 
in order to evaluate its compliance with the ASTRI SST-2M focal plane requirements.
In particular, we investigated the trigger time walk and the jitter effects as a function 
of the pulse amplitude.
The EASIROC output signal is found to vary linearly as a function of the input pulse 
amplitude with very low level of electronic noise and cross-talk ($<$1\%).
Our results show that it is suitable as front-end chip for the camera prototype, although, 
specific modifications are necessary to adopt the device in the final version 
of the telescope.

\end{abstract}


\end{frontmatter}

\section{Introduction}
The use of Cherenkov telescopes in gamma-ray astronomy opened a new frontier in the 
study of the emission from very energetic sources such as supernovae, 
neutron stars and supermassive black-holes.
Since the discovery of TeV emission from the Crab nebula with Whipple 
\cite{weeks89} in 1989, the number of known sources emitting in this range 
of energy has rapidly increased to hundreds in a relatively small number 
of years.
The Cherenkov Telescope Array (CTA) is a new generation of ground based 
gamma-ray telescopes aiming at observing the sky from a few tens of GeV to beyond 100 TeV.
It is an array comprising telescopes with three different sizes 
each designed to cover a specific range of energy \cite{cta10}. 
The low energy band (from  a few tens of GeV up to a hundred GeV)  
will be covered by 24 m mirror diameter telescopes that observe 
a Field of View (FoV) of 4$^\circ$-5$^\circ$.
The range 100 GeV-1 TeV will be observed by 10-12 m mirror diameter telescopes having
 a FoV of 6$^\circ$-8$^\circ$. 
The highest energy region will be observed by small size telescopes 
(4-6 m mirror diameter) having a FoV of about 10 degrees.\\
ASTRI (Astrofisica con Specchi a Tecnologia Replicante Italiana; \cite{canestrari11}), 
is a 'Flagship Project' of the Italian Ministry of Education, University and
Research led by the Italian Institute of Astrophysics (INAF), 
whose main object is to build a prototype for the small size CTA telescopes. 
The prototype, hereafter named ASTRI SST-2M telescope, will adopt innovative solutions 
to explore the 1-100 TeV range of the electromagnetic spectrum: a wide field dual 
mirror Schwarzschild-Couder optical system and 
Silicon Photo-Multipliers (SiPM) as sensors of the camera at the focal plane. 

The telescope design foresees a focal ratio F\# = 0.5, an equivalent focal length of 2150mm,
an average effective area of about 6.5m$^2$, and, considering the size of the focal plane 
surface, it covers a Field of View (FoV) of $\sim$9.6$^\circ$ in diameter.

 The detection of Very High Energy (VHE) gamma photons is possible by observing the 
Cherenkov light produced by the relativistic secondary particles
in the air shower generated by the interaction with the Earth atmosphere. 
 This light, emitted in the ultraviolet band within a cone of about 1.3$^\circ$, 
reaches the ground in a '' pool '' having a radius of $\sim$120 m.
 It is very faint and lasts only a few ns with a duration that depends mainly
on the distance between the shower core and the telescope axis \cite{heb11}.

SiPMs are suitable for the detection of the Cherenkov flashes,
as already  demonstrated by the FACT project \cite{anderhub09}, because they are
very fast and sensitive to the light in the range 300-900 nm. 
Their drawbacks with respect to the traditional PMTs (PhotoMultiplier Tubes) are: 
a very high dark count, after pulses, optical cross-talk between  elementary diodes inside a pixel
and a gain strongly dependent on chip temperature. Nevertheless, for a single pixel 
the rate of the dark counts is lower than the rate of the Night Sky Background (NSB), so that the high 
instrumental background does not degrade the telescope sensitivity.
Moreover, the effects of optical cross-talk and after pulses, that may affect the sensitivity, 
are typically lower than 20\%\cite{eckert10} and the gain can be kept stable with an adequate strategy of 
 temperature control.

In this paper, we present a set of measurements  aimed at
characterizing the performance of EASIROC, a device devoted to read the SiPM output, 
and compare the results  with ASTRI SST-2M focal plane requirements.

\section{ The ASTRI Camera}
\begin{figure*}[ht!]
\centering
\includegraphics[angle=0, width=5cm]{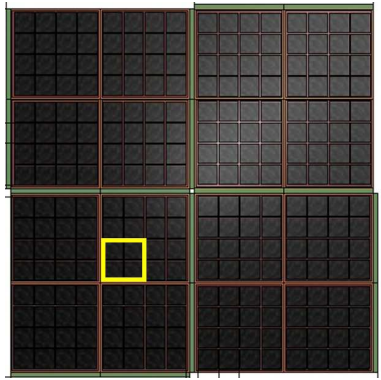}
\caption{A PDM of ASTRI SST-2M focal plane composed of 4$\times$4 SiPMs. 
The yellow square shows the logical pixels obtained grouping 2$\times$2 physical pixels}
\label{fig1}
\end{figure*}

The camera at the focal plane of the ASTRI SST-2M is based on  monolithic SiPMs 
Hamamatsu~S11828-3344m\footnote{http://www.hamamatsu.com/sp/hpe/HamamatsuNews/HEN111.pdf}
with 4x4 squared pixels, 3$\times$3mm large, made up
of 3600 elementary diodes of 50$\mu$m pitch giving a filling factor of 62\%.
The physical pixels are 2$\times$2 grouped in logical pixels (6.2$\times$6.2mm)
shown with a yellow square in Fig.~\ref{fig1}. 
Therefore, considering that, at the focal plane, 1$^\circ$ corresponds to 37.5mm, 
the logical pixel has a size of 0.17$^\circ$ matching the optics angular resolution.
The detector units are organized in an array of 37 Photon Detection Modules
(PDM) with 8$\times$8 logical pixels each (see Fig.~\ref{fig1}), 
for modularity and a fast read out of the focal plane
after any trigger. The trigger threshold is chosen to 
have a maximum rate of $\sim$300 Hz from the whole focal plane that ensures a dead time
$<$3\%.
 The energy working range of the the ASTRI SST-2M telescope prototype 
 is 1-100 TeV and, considering 
the optics area, the requirements for the maximum number of photoelectrons 
detected in one pixel is 1000 with a goal of 2000.

The very short duration of gamma events requires  a dedicated FEE (Front-End Electronics) capable,
not only to catch the very fast pulses of Cherenkov light, but also to provide auto-trigger
capability.
The device proposed for the ASTRI SST-2M telescope prototype is the Extended Analogue
Silicon Photo-Multiplier Integrated Read Out Chip 
(EASIROC; \cite{callier11}) equipped with 32-channels each with the capability of measuring charge 
from 1 to 2000 photoelectrons assuming a SiPM gain of 10$^6$. Two EASIROC devices are then
devoted to read a single PDM.

\section{The EASIROC chip}
\begin{landscape}

\begin{figure*}[!ht]

\includegraphics[angle=0, width=19cm,height=12cm]{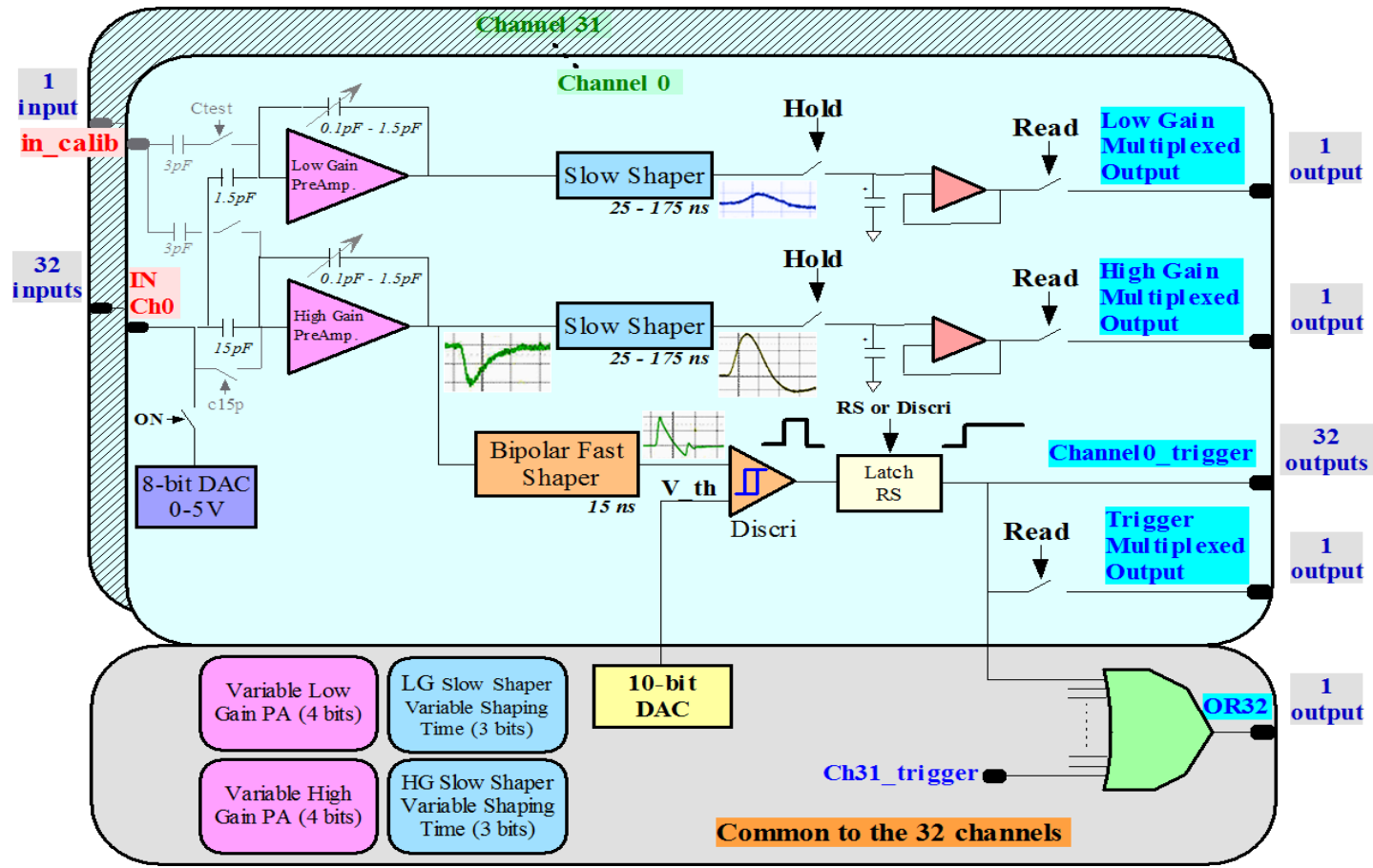}
\caption{Architecture of the front-end EASIROC (Omega Micro courtesy)}
\label{fig2}
\end{figure*}

\end{landscape}

 EASIROC, produced by Omega Micro\footnote{http://omega.in2p3.fr}, is a 32 channel
 fully analogue front-end  Application-Specific Integrated Circuit
(ASIC) dedicated to the readout of SiPM detectors.
The architecture of its analog core and the characteristics are shown 
in Fig.~\ref{fig2} and Table~\ref{table1}, respectively.

In each channel, two separate electronics chains allow High- and Low-gain (HG/LG) 
of the signal in order to measure  charge from 160 fC up to 320 pC 
covering a range up to 2000 photoelectrons for the adopted $10^6$ SiPM gain.

The high voltage of each logical pixel can be individually tuned by varying 
the bias voltage in a 8-bit DAC from 0 to 4.5V in order to equalize the gains.

Each of the two chains is composed of an adjustable preamplifier followed by a tunable 
shaper and a track and hold circuit.
A third chain is implemented to generate a trigger using a fast shaper followed by a discriminator, 
whose threshold is set by a 10-bit DAC common to all the 32 channels (see Fig.~\ref{fig2}).
The signal shaping time is set to 50 ns 
according to our needs, while the trigger shaping time is fixed to 15 ns.

The power consumption is lower than 5 mW/channel and unused features can be disabled 
to further save power.
All EASIROC main parameters can be programmed by downloading a configuration table
through a slow control serial line.
The  processing of the analog signal takes place in the front-end channels of the device, 
while the read-out is handled at the internal back-end of the ASIC.
The outputs of all the channels can be readout from multiplexers that, running in parallel, 
can  sequentially switch the sampled signal of the two chains (LG and HG).

\begin{figure*}[ht!]
\centering
\includegraphics[angle=0, width=10cm]{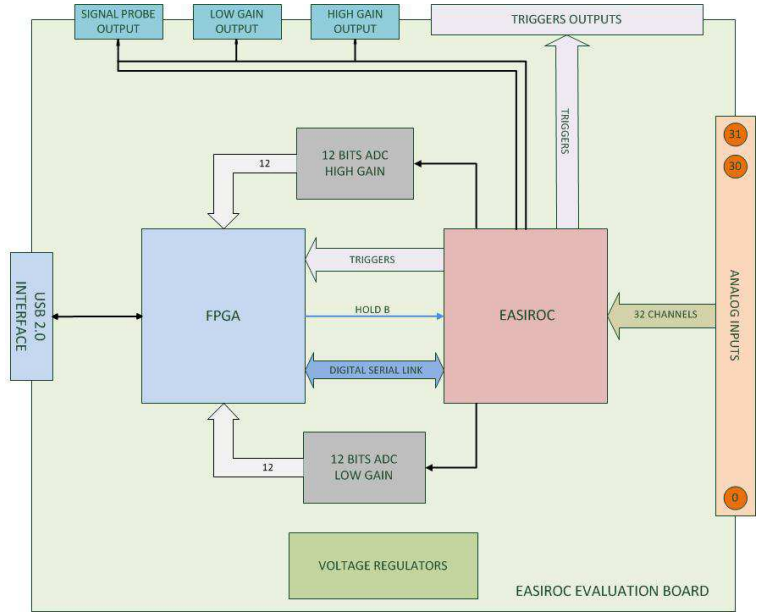}
\caption{Architecture of the PCB (Printed Circuit Board) of the EASIROC (Omega Micro courtesy)}
\label{fig3}
\end{figure*}

An evaluation board has been designed and realized by Omega Micro
to test the functional characteristics and performance of this ASIC. 
It allows an easy access to the EASIROC output and  provides many test points 
to the Field Programmable Gate Array (Altera Corporation - Cyclone FPGA Model EP1C6Q240C6N)
 dedicated lines (see Fig.~\ref{fig3}). 
It is equipped with two external Analog-to-Digital Converters (ADC) to allow ASIC digital data acquisition.  
The trigger of each channel and the OR output of the 32 trigger channels ($OR32$) 
are available in parallel.
The 32 triggers are sent to the FPGA that, according to the set threshold, generates 
a common hold signal ($HOLD-B$) for the output read-out.
 A LabVIEW software of the National Instruments, developed by the LAL 
(Laboratoire de l'Acc$\acute{e}$l$\acute{e}$rateur Lin$\acute{e}$aire) Tests group\footnote{ http://www.lal.in2p3.fr/},
has been provided together with the evaluation board to command the EASIROC 
chip and to receive the output via Universal Serial Bus (USB) connection. 
It allows one to send the ASIC configuration  and to receive the output bits via a USB cable connected to the
test board. 

In our measurements, we used a new version of the chip, 
EASIROC-A, kindly provided by Omega Micro.

\setcounter{table}{0}
\begin{table*}[htbp!!]
\centering

\caption{Main characteristics of the EASIROC chip.}
\label{table1}
\begin{tabular}{|l|l|}
\hline
Technology:     & Austria-Micro-Systems (AMS) SiGe 0.35 $\mu$m\\
\hline
Dimensions :    &    16.6 mm$^2$ (4.15$\times$4.01mm)  \\
 \hline
Power Supply : & 4.5V/0V \\
 \hline
Consumption:    & 4.84mW per channel \\
\hline
Package :       & Naked (PEBS) TQFP160  \\
\hline
 \end{tabular}
 \end{table*}

 \newpage

\section{Laboratory set-up}

A pulse function arbitrary generator is used to create the input charge
injected in EASIROC. The signal is similar, as much as
possible, to the SiPM one characterized by very fast rise time (a few hundreds 
of ps) followed by an exponential decay a few tens of ns).
 The amplitude of this signal for an input charge 
of 0.16 pC, equivalent to 1 pe for a SiPM gain of 10$^6$, is 300$\pm$4$\mu$V.

According to the goal on the ASTRI SST-2M maximum rate per pixel, 
 the gain of the LG chain preamplifier is set to 3 
that produces a monotonic response up to 2000 pe.
Moreover, we decided to fix the gain of the HG chain preamplifier to 150
allowing an almost linear working range up to $\sim$50 pe (see Sect. 6.3).

 The shaping time of the signal was fixed to 50 ns, considering that the higher energy events, 
mainly coming from large shower core distance, are expected to last up to $\sim$30 ns.

In our sets of measurements, we first characterised the trigger behaviour 
and the delays introduced by the jitter and by the time walk. 
We then sampled the shape of the output signal for both HG and LG chains
and investigated  its linearity with respect to the injected charge.
Finally, we evaluated the level of the cross-talk. 
All the measurements were performed using only one of the 32 channels  ({\it Ch~31}).  
However, to check the validity of the results, some measurements were repeated 
for other channels.

\section{Trigger}

The trigger chain, derived from the HG preamplifier, is composed of a
dedicated 15 ns fast shaper followed by a discriminator which provides the trigger signal. 
The threshold of the discriminator is common to all the 32 channels and it is set by a 10-bit 
DAC whose output range goes from 1.06V up to 2.38V, in steps of 1.28mV.
 
To characterise the DAC-controlled discriminator, we tested the threshold voltage at the 1024 DAC values
using a Keithley 2000 Digital Multimeter and we found that Voltages follow a straight
 line with an average discrepancy of $\sim$0.5$\%$.

\begin{figure*}[!h]
\centering
\includegraphics[angle=0, width=8.5cm]{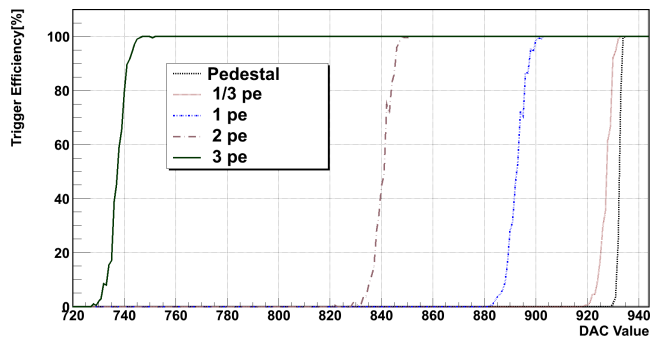}
\caption{ Trigger efficiency for fixed injected charges as function of the discriminator threshold.
The dotted line at 930 DAC counts corresponds to the trigger efficiency with no injection charge}
\label{fig4}
\end{figure*}

 The trigger output was evaluated varying the discriminator threshold for fixed  
injected charge in the range 0-8 pC (0-50 pe) using a reference clock (10kHz) synchronous to the signal.
As an example, we show in Fig.~\ref{fig4} the efficiency curves for 0.33 pe, 1 pe, 2 pe and 3 pe.
The last dotted line at $\sim$930 DAC corresponds to the trigger efficiency with 
no injection charge, obtained keeping the same reference clock. 
In this case, triggers are due to the electronics noise.

\begin{figure*}[!h]
\centering
\includegraphics[angle=0, width=8.5cm]{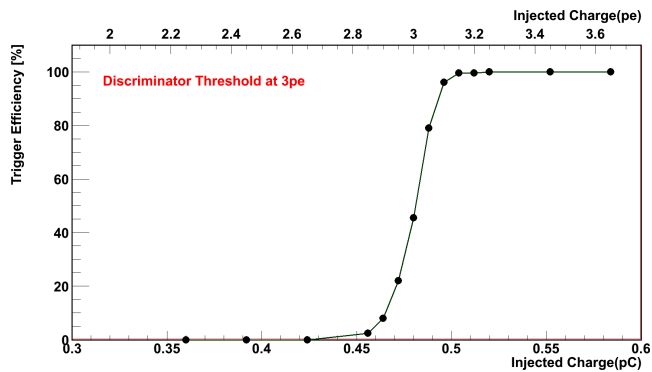}
\caption{Trigger efficiency vs injected charge setting the discriminator threshold to 3 pe.}
\label{fig5}
\end{figure*}

From simulations, we evaluated that the requirement of 300Hz
on the total trigger rate at the focal plane  is ensured
if at least four contiguous pixels are triggered, with the threshold set at 3 pe.
The trigger efficiency as function of the injected charge is shown in Fig.~\ref{fig5} 
for a threshold set to 3 pe.

\newpage

\subsection{Time Walk and Jitter}

The time walk and jitter of the trigger were investigated by measuring, at the oscilloscope,
the delay between the trigger at the chip and the strobe of the generator.  
The range of injected charge is 0.16--3.2 pC (1-20 pe) and
the discriminator threshold level is sequentially set to 0 pe (the pedestal), 1 pe, 2 pe and 3 pe.
\\
Since these measurements could be influenced by the  jitter in the signal, 
we first investigated  its level and found that it is
negligible being of the order of a few tens of ps.

\begin{figure}[!h]
\centering
\includegraphics[width=8.5cm]{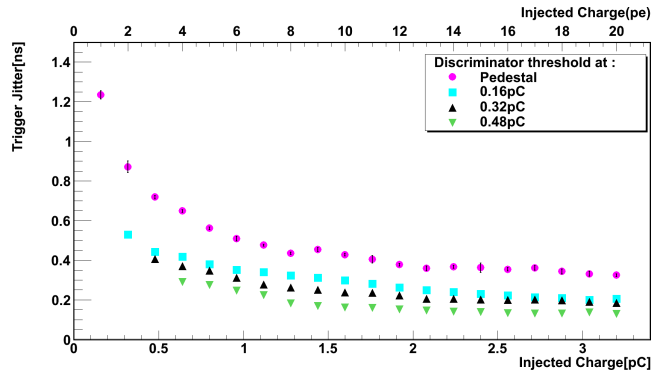}
\caption{Jitter of the trigger vs injected charge at several discriminator thresholds.}
\label{fig6}
\end{figure}

 The time jitter, measured as the RMS of the delay distribution, are presented 
in Fig.~\ref{fig6} as a function of injected charge.
As expected, it is higher at pedestal but, in the sampled charge range,
its value is always lower than 1.23 ns and decreases with increasing threshold.
For a discriminator threshold of 3 pe and injected charges higher than 0.64 pC (4 pe) it is lower than 0.3 ns. 
 
\begin{figure}[!h]
\centering
\includegraphics[width=8.5cm]{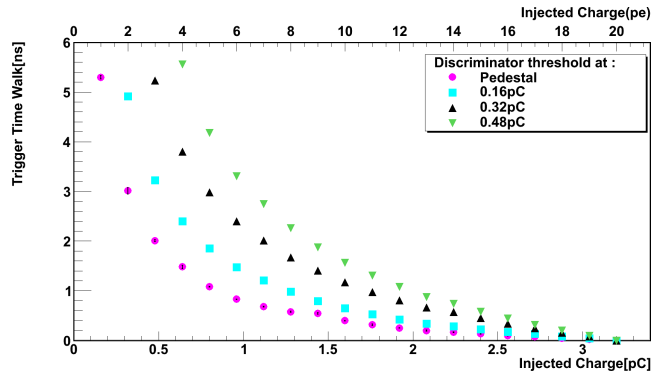}
\caption{Time walk of the trigger vs injected charge at several discriminator thresholds.
The zero level was set to the time walk corresponding to the largest injected charge.}
\label{fig7}
\end{figure}

We show in Fig.~\ref{fig7} the values of the time walk as function of injected charges. 
We measured them as the mean of the delay distribution
and shifted them so that the time walk corresponding to the largest injected charge is zero in each case. 
The time walk is always below 5.5 ns and it is reduced to 1 ns for signals 
higher than 1.92 pC (12 pe) for a discriminator threshold of 3 pe.

\newpage
\section{Output signal}
The expected range of the Cherenkov signal (1-2000 pe) requires the characterization of 
both HG and LG chain.
Our measurements investigated the level of the electronic noise for different working 
conditions of the chip, the shape of the output signal and the linearity as function 
of the injected charge.

\subsection{Electronic Noise}

The level of the noise affecting the signal was evaluated as function of the preamplifier gain.
We defined the noise as the $RMS$ of the analogue output signal with no charge injected. 
To convert the voltages into Equivalent Noise Charge (ENC),
we used the following formula:
\begin{equation}
\label{fm1}
 ENC\,=\,K\times V_{RMS} \hspace{0.5cm} 
{\rm with}\hspace{0.5cm}K= \frac {Q _{N pe  }}{V_{N pe  }}
\end{equation} 
\noindent
where $V_{RMS}$ is the rms spread in output voltage due to the noise, ${Q _{N pe   }}$ and
${V_{N pe  }}$ are the charge and the output voltage corresponding 
to a signal $N pe$, respectively. We found that $K = 0.045$ [pe mV$^{-1}$] 
for gain = 150 in the HG and $K = 1.15$ [pe mV$^{-1}$]
for gain = 3 in the LG chain.   

\begin{figure*}[!t]
\centering
\includegraphics[angle=0, width=12cm]{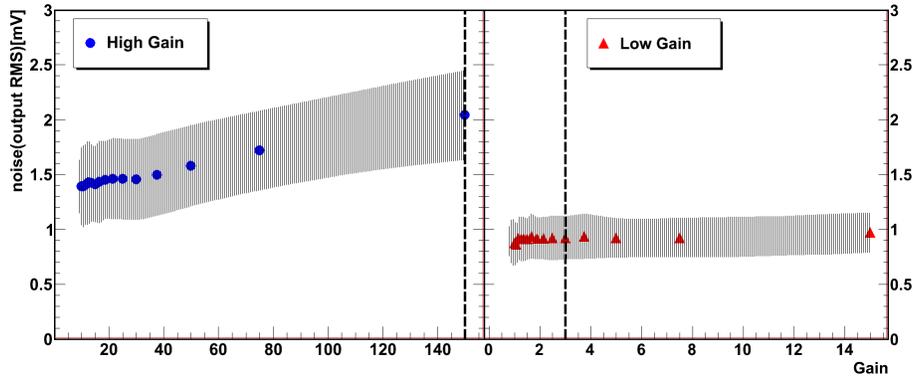}
\caption{ Average electronic noise in mV measured at the oscilloscope vs the preamplifier 
gain. The vertical dashed lines indicate the gains at which the EASIROC will be operated.
 The shaded areas represent the noise variability ranges.
 The left panel relates to the HG and the right panel to the LG.}
\label{fig8}
\end{figure*}

The feedback capacitance (see Fig.~\ref{fig2}) was varied in its working range
0.1-1.5 pF corresponding to the gain range 150-10 in the HG chain 
and to 15--1 in the LG one. The average noise together with its variability range 
are shown in Fig.~\ref{fig8}, for both gains.
In the HG chain, the noise is 2.1 mV at gain =150, which
converts to 0.1 pe using the Equation~\ref{fm1}. 
\\
The noise in the LG chain  is constant for all the investigated
gains and its average value is 0.91$\pm$0.05 mV, equivalent to $\sim$1 pe.

\newpage
\subsection{Slow-shaper sampling}
The chip saves the amplitude of the preamplified and shaped signal at 
a given time, the hold time,  directly set by the user using a track and hold cell 
(see Fig.~\ref{fig2}). 
Two external ADCs, one for the HG and one for the LG, digitize the signal level measured 
at the hold time (see Fig.~\ref{fig2}).

We sampled the shaped signal for the two electronics chains at several
hold times with the main task of evaluating the peaking time.
Each point was derived averaging over 1000 samplings and
several different injected charges were investigated fixing the shaping time
to 50 ns.

\begin{figure*}[!ht]
\centering
\includegraphics[width=8.5cm]{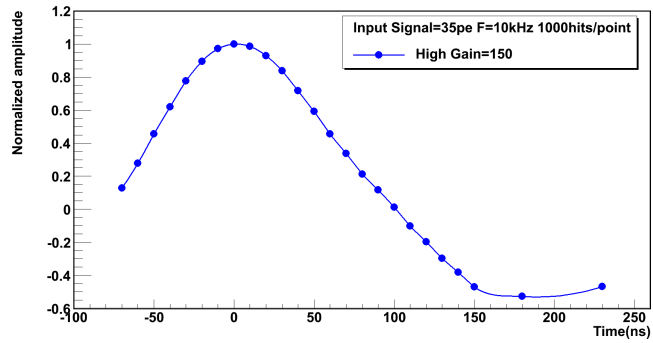}
\caption{\label{fig9} Amplitude normalized to the ADC value at the peaking time
 versus the time in ns shifted at the peaking time  for the HG chain.}
\end{figure*}

\begin{figure*}[!h]
\centering
\includegraphics[width=8.5cm]{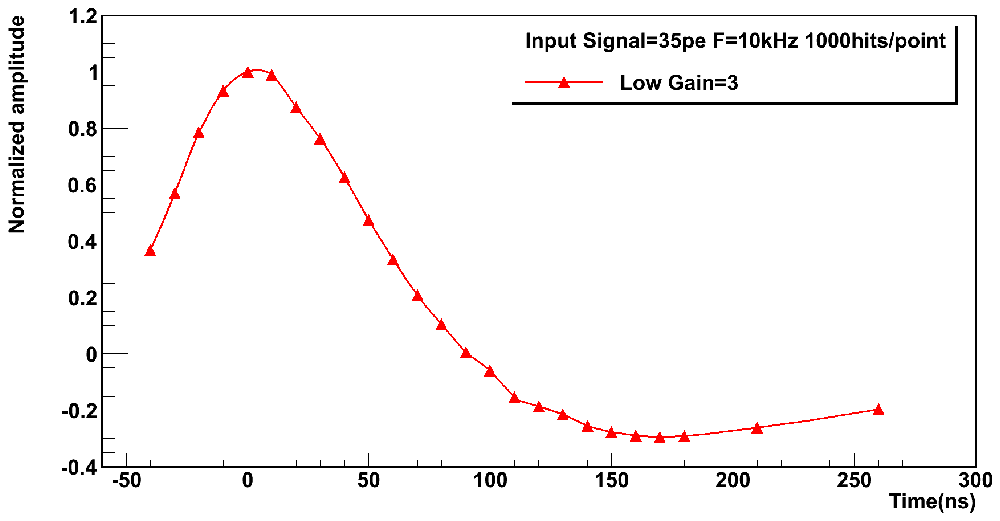}
\caption{\label{fig10}Amplitude normalized to the ADC value at the peaking time
 versus the time in ns shifted at the peaking time  for the LG chain.}
\end{figure*}

\begin{figure*}[ht]
\centering
\includegraphics[width=8.5cm]{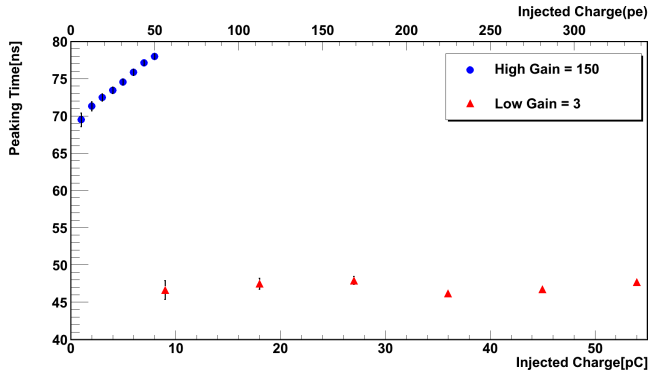}
\caption{\label{fig11}Peaking time vs the injected charge. Red points are relate to the LG chain, blue points
to the HG one. }
\end{figure*}

The pedestal-subtracted signal was normalized to give a peak amplitude of one, 
and the time axis was adjusted so that this peak was centred at 0. 
The curves as a function of time in ns are shown in ~\ref{fig9} and ~\ref{fig10} for the HG and LG chains,
 respectively. We note that, 10 ns after the peaking time, the amplitude of the signal produced by 
 an injected charge of 5.6 pC is reduced by $\sim$25 ADC counts with respect to the peak 
(corresponding to $\sim$1.4\% in the normalized curve of Fig.~\ref{fig9}) 
for the HG chain, and by $\sim$1 ADC counts (corresponding to $\sim$1\% in the curve of Fig.~\ref{fig10}) 
for the LG chain.
\\
 In order to compute the peaking time, the peak of the signal
was fitted with a Gaussian whose best fit values 
are presented in  Fig.~\ref{fig11} for the two gain chains.
We note that, the peaking time of the LG is quite constant: it
varies by 0.6 ns with respect to its average (47.1 ns) 
over the entire investigated range (56-338 pe).
The case is different in the 6-50 pe range of the HG chain where the peaking time varies  by
about 6\% with respect to the average of 74.0 ns that is moreover 24 ns higher than the designed 
shaping time (50 ns).

\clearpage

\subsection{Charge linearity}

\begin{figure*}[!ht]

\centering
\includegraphics[width=8.5cm]{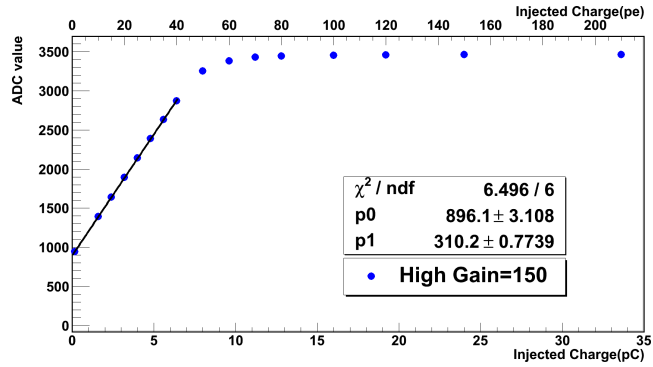}
\caption{\label{fig12}ADC values vs injected charges for the HG chain. 
The continuous line is the linear best fit up to the injected charge of 6.4 pC (see text).}
\end{figure*}

\begin{figure*}[!h]
\centering
\includegraphics[width=8.5cm]{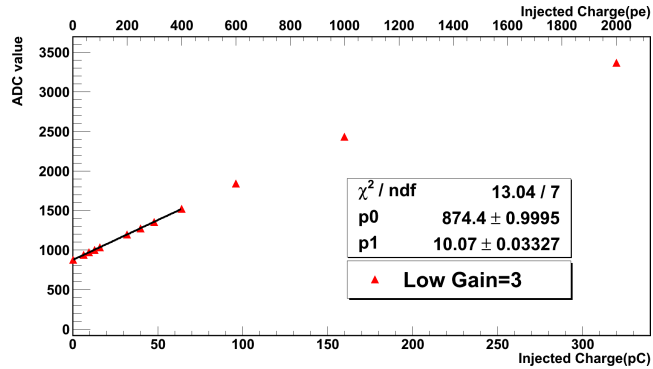}
\caption{\label{fig13}ADC values vs injected charges for the LG chain.
The continuous line is the linear best fit up to the injected charge of 64 pC (see text).}

\end{figure*}

The linearity of the system was checked by analyzing the slow shaper amplitude
at the time 50 ns for the LG chain and 80 ns for the HG one 
as function of the injected charge.  We assumed as error of each point the sigma 
of the distributions obtained performing 1000 tests. 
These uncertainties are always lower than 1\%. Furthermore, the Integral and Differential Nonlinearity specified
for the device AD9220 is much smaller than this uncertainty\footnote{http://www.alldatasheet.com/datasheet-pdf/pdf/237229/AD/AD9220.html}.  
The ADC values as function of injected charges are shown in Fig.~\ref{fig12} and Fig.~\ref{fig13}
for the LG and the HG.

 We note that the ADC values curve has a monotonic behaviour up to 8 pC (50 pe) 
for the HG and up to 320 pC (2000 pe) for the LG. 
These two working ranges are then suitable for the ASTRI SST-2M requirements (see Sect.4).
  
Moreover, we established the range where the correlation between the ADC values and the charge 
can be modelled with a straight line.
Starting from a few points we increased the fitting range up to the maximum charge 
that still gives an acceptable $\chi^{2}$.
We find that the HG chain is linear in the range 0.16-6.4 pC and the LG chain in the range 0.16-64 pC,
where $\chi^{2}$ =6.5 for 6 degrees of freedom(dof) and $\chi^{2}$ =13 (7 dof)
were obtained for HG and LG, respectively. The measured points and the fitting lines in the validity ranges
 are shown in Fig.~\ref{fig12} and Fig.~\ref{fig13}.

\newpage

 \section{Cross-Talk}

Our last step of measurements were devoted to investigate
 the cross-talk as function of the injected charge.
 This  was performed measuring the output signal in the two $Ch~31$ neighboring channels  
 ($Ch~0$ and $Ch~30$) for input charges ranging from 1 to 320 pC (6-2000 pe).
We assumed that a cross-talk signal is detected if its intensity is 2.5 $\sigma$ 
above the noise.
We find that no significant cross-talk was observed for the LG  
in the whole investigated range and for the HG below 5 pC (31 pe).

\begin{figure*}[!ht]
\centering
\includegraphics[width=8.5cm]{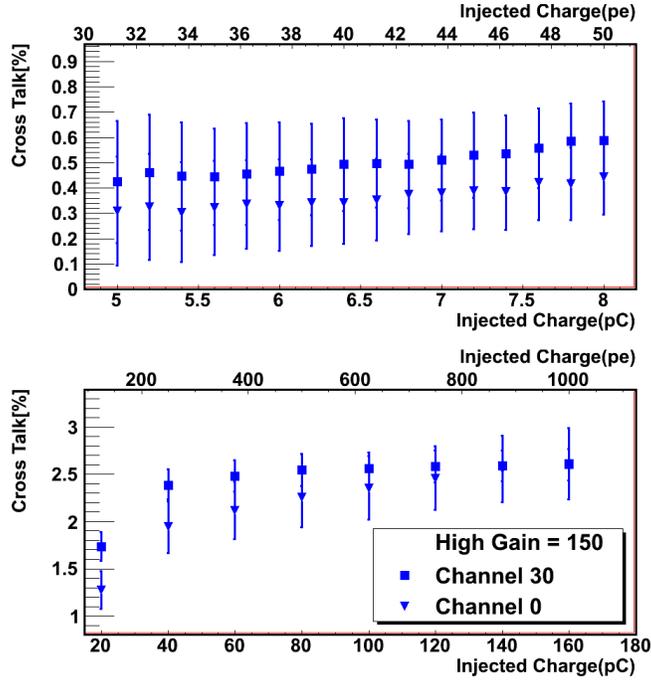}
\caption{\label{fig14}The ratios, in percentage, between the outputs in $Ch~0$ and $Ch~30$ and the output
in $Ch~31$ vs injected charge for the HG chain. 
The top panel corresponds to the ASTRI working range ($<$50 pe);
the bottom panel shows the cross talk in the range where the HG chain saturates.}
\end{figure*}

The ratios, in percentage, between the outputs in $Ch~0$ and $Ch~30$ and the output
in $Ch~31$ for the HG chain for input charges above 5 pC are presented in Fig.~\ref{fig14}.
We note that the effect of the cross-talk is always lower than 1\%
in the ASTRI SST-2M HG working range ($<$50 pe) and its level is  
$<$3\% in the range where HG saturates (bottom panel of Fig.~\ref{fig14}).
 
These values must be compared with the cross-talk between logical pixels, which 
is not known at the moment. Dedicated measurements are foreseen in the SiPM characterization.

 \newpage

\section{Summary and Conclusion}
We performed a set of measurements to investigate the performance of EASIROC as the ASIC
front-end for the camera at the focal plane of the ASTRI SST-2M telescope prototype. 
The measurements were mainly aimed at understanding the
triggering capability and the shape of the output signals.

\begin{table*}[htbp!!]
\centering

\caption{Experimental results of EASIROC characterization.}
\label{table2}
\begin{tabular}{|l|l|}
\hline
Jitter(discriminator thresholds 3 pe)    & 0.3 ns\\
\hline
Time walk(discriminator thresholds 3 pe) & $<$5.5ns  \\
 \hline
 Noise HG     &   $\sim$0.1 pe \\
 \hline
  Noise LG     &   $\sim$1 pe  \\
 \hline
Cross Talk HG &   $<$1$\%$      \\
 \hline
Cross Talk LG   & No            \\
\hline
 \end{tabular}
 \end{table*}

The main experimental results are presented in the Table~\ref{table2}.
They are summarized in the following points and compared with the ASTRI SST-2M  
requirements:
\begin{itemize}
\item
{\it Trigger:} EASIROC is able to trigger events with a number of counts $>$0.3 pe, 
well below the value foreseen for the ASTRI SST-2M prototype. In fact, the requirement on the maximum
trigger rate (300 Hz) can be obtained if at least four contiguous pixels 
each have a signal greater than 3 pe. \\
The trigger time walk and jitter are always below 5.5 ns and 0.3 ns, respectively, 
for a trigger threshold set to 3 pe.  
They introduce an uncertainty in the charge
measurements of a few percent in both chains, with the 
shaping function of  EASIROC configured to give a signal with a broad peak.

\item
{\it Signal:} 
The output signal has a stable peaking time in the LG chain and its value corresponds to the
set shaping time. Some problems are detected in the HG chain. The peaking time is on average shifted
by $\sim$20 ns with respect to the shaping time with a significant drift
as function of the injected charge. 
Since the hold signal $HOLD-B$ is common to both the chains in the present version 
of the chip, a reduction in the amplitude of the HG signal is measured (see Fig.\ref{fig9}). 
This effect can be reduced, without degrading the performance of the LG chain,  
 by choosing a suitable sampling time. 
However, a new version of the chip should correct this shortcoming either by anticipating the 
HG peaking time or by providing two different hold signals.

We detected a monotonic behaviour up to 8 pC for HG chain
and up to 320 pC for the LG one. 
The levels of the electronic noise and the cross-talk between channels introduced
by EASIROC are negligible.

\end{itemize}

\noindent
In conclusion, our measurements showed that EASIROC is suitable for the front-end of 
the camera at the focal plane of the ASTRI SST-2M telescope prototype. 
Minor ad-hoc modifications in its re-design, as for example a different $HOLD$ signal
for each electronics chain, are however necessary to tailor this device to be fully 
compliant with the telescope requirements.

\section*{Acknowledgements}
The work presented in this paper was partially supported by the ASTRI,
"Flagship Project" financed by the Italian Ministry of Education, University,
and Research (MIUR) and lead by the Italian National Institute of Astrophysics (INAF).
We are deeply grateful to S. Callier, C. De La Taille,
and L. Raux of the Omega Micro at Orsay and to M.C. Maccarone, A. Rubini and G. Tosti of the
ASTRI collaboration for useful discussions and suggestions.







\bibliographystyle{model1c-num-names}
\bibliography{<your-bib-database>}







\end{document}